\documentclass[twocolumn, showpacs, showkeys, amsmath, amssymb, superscriptaddress, prb, citeautoscript]{revtex4-1}

\usepackage{amsmath,amsfonts,amssymb}
\usepackage{graphicx}			
\usepackage{dsfont}  			

\usepackage[normalem]{ulem} 	
\usepackage[usenames,dvipsnames]{color}

\usepackage[normalem]{ulem} 	

\newcommand{\moved}[1]{\sloppy{\textcolor{blue}{\sout{#1}}}}
\newcommand{\delete}[1]{\sloppy{\textcolor{red}{\sout{#1}}}}

\newcommand{\mcom}[1]{\marginpar{#1}} 

\newcommand{\ket}[1]{\ensuremath{\left|#1\right\rangle}}
\newcommand{\mean}[1]{\ensuremath{\left\langle #1\right\rangle}}
\newcommand{\nn}{\nonumber}

\renewcommand{\delete}[1]{}

\renewcommand{\mcom}[1]{}
\renewcommand{\moved}[1]{}

\renewcommand{\figurename}{Figure} 

\begin{document}

\title{Observation of directly interacting coherent two-level systems in a solid}

\author{J\"urgen Lisenfeld}
	\affiliation{Physikalisches Institut, Karlsruhe Institute of Technology,
76131 Karlsruhe, 
	Germany}
\author{Grigorij J. Grabovskij}
	\affiliation{Physikalisches Institut, Karlsruhe Institute of Technology,
76131 Karlsruhe, 
	Germany}
\author{Clemens M\"uller}
	\affiliation{D\'epartement de Physique, Universit\'e de Sherbrooke, Sherbrooke, Qu\'ebec, 	Canada J1K 2R1}
	\affiliation{ARC Centre of Excellence for Engineered Quantum Systems, School of Mathematics and Physics, University of Queensland, Brisbane, Queensland 4072, Australia}
\author{Jared H. Cole}
	\affiliation{Chemical and Quantum Physics, School of Applied Sciences,
RMIT University, 
	Melbourne, 3001, Australia}
\author{Georg Weiss}
	\affiliation{Physikalisches Institut, Karlsruhe Institute of Technology,
76131 Karlsruhe, 
	Germany}
\author{Alexey V. Ustinov}
	\affiliation{Physikalisches Institut, Karlsruhe Institute of Technology,
76131 Karlsruhe, Germany}
\affiliation{Russian Quantum Center, 100 Novaya St., Skolkovo, Moscow region 143025, Russia}

\date{\today}
\begin{abstract}
Parasitic two-level tunneling systems originating from structural material defects affect the functionality of various microfabricated devices by acting as a source of noise. In particular, superconducting quantum bits may be sensitive to even single defects when these reside in the tunnel barrier of the qubit's Josephson junctions, and this can be exploited to observe and manipulate the quantum states of individual tunneling systems.

Here, we detect and fully characterize a system of two strongly interacting defects using a novel technique for high-resolution spectroscopy. Mutual defect coupling has been conjectured to explain various anomalies of glasses, and was recently suggested as the origin of low frequency noise in superconducting devices. Our study provides conclusive evidence of defect interactions with full access to the individual constituents, demonstrating the potential of superconducting qubits for studying material defects. All our observations are consistent with the assumption that defects are generated by atomic tunneling.

\end{abstract}
\maketitle
\section*{Introduction}
Superconducting quantum bits~\cite{ClarkeWilhelmReview} have recently achieved a breakthrough by demonstrating excellent gate fidelities and long coherence times in a fully scalable architecture~\cite{Barends14}, placing the realization of an integrated quantum computing chip within reach. The solid-state approach however bears the burden that the material of the quantum device itself may host parasitic defects that give rise to two-level systems (TLS) acting as a sparse decohering bath. 

First signatures of coherent TLS in phase qubits were found in spectroscopy data, where observed avoided level crossings manifest the defects' two-level quantum character~\cite{Simmonds:PRL:2004,Martinis:PRL:2005}. 
Often, these defects show longer coherence times than the qubit itself~\cite{Lisenfeld:PRL:2010} and thus might be useful as quantum memories~\cite{Neeley:NP:2008} and resources for quantum
algorithms~\cite{Muller:PRA:2012, Grabovskij:NJP:2011}. 
Phase qubits were employed in several attempts to identify the physical origin of those TLS, for example by obtaining statistics on frequencies and coupling strengths~\cite{Shalibo:PRL:2010}, estimating their density~\cite{Martinis:PRL:2005}, measuring the temperature dependence of their coherence times~\cite{Lisenfeld:PRL:2010}, or verifying theoretical models describing their origin~\cite{Cole:APL:2010}. 

The possibility of a direct interaction between TLS has been invoked in the past to explain the line width broadening and spectral diffusion of ultrasonically excited ensembles of TLS in glasses~\cite{ArnoldHunklinger75,Black:PRB:1977} as well as various other low-temperature properties of disordered solids\cite{HuEnss2000,esquinazi}. TLS are furthermore a widely accepted model to explain noise in superconducting circuits, and mutual TLS coupling was recently suggested as the origin of the low frequency noise observed in microwave resonators~\cite{FaoroResonators}. \\[0.4cm]

\begin{figure}[htb!]
\centering\includegraphics{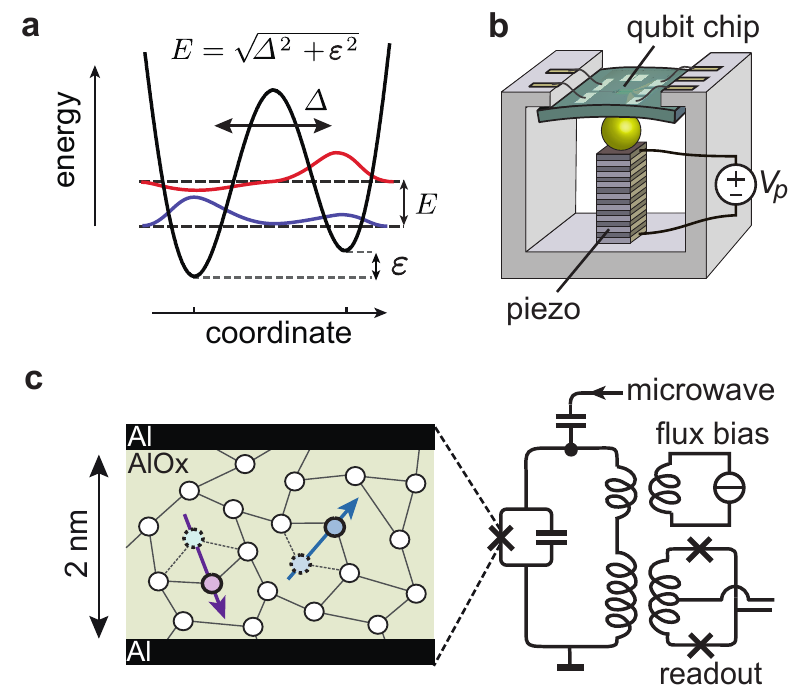}
\caption{\label{fig:figure1}\textbf{Using a superconducting qubit to access defects in Josephson junctions.} \textbf{a}  Illustration of the double-well potential for an atomic tunneling system. Tunneling energy $\it\Delta$ and asymmetry energy $\varepsilon$ determine the level splitting $E$.
\textbf{b} Sketch of the sample holder. To control the strain, the qubit chip is bent by applying a voltage $V_\mathrm{p}$ to the stacked piezo actuator.
\textbf{c} Schematic of the phase qubit including manipulation and measurement circuitry. The Josephson junction (JJ) tunnel barrier is sketched as a disordered insulator hosting TLS defects, here pictured as atoms tunneling between two metastable positions, with the arrows illustrating their electric dipole moment.
}
\end{figure}

Here, we report the first clear experimental evidence of two coherently interacting  two-level systems which are residing in the tunnel barrier of a Josephson junction. The data are obtained with a new technique for high-resolution defect spectroscopy that exploits the tunability of TLS by mechanical strain and their strong coupling to a superconducting qubit. To characterize the coupled defect system in more detail, we extend upon this technique and perform coherent two-photon spectroscopy that directly reveals the TLS' coupling strengths and independent parameters.
Interpretation of the measurement based on atomic tunneling systems fully accounts for all observations.

\begin{figure*}
	\centering\includegraphics[width=18.3cm, height=8cm]{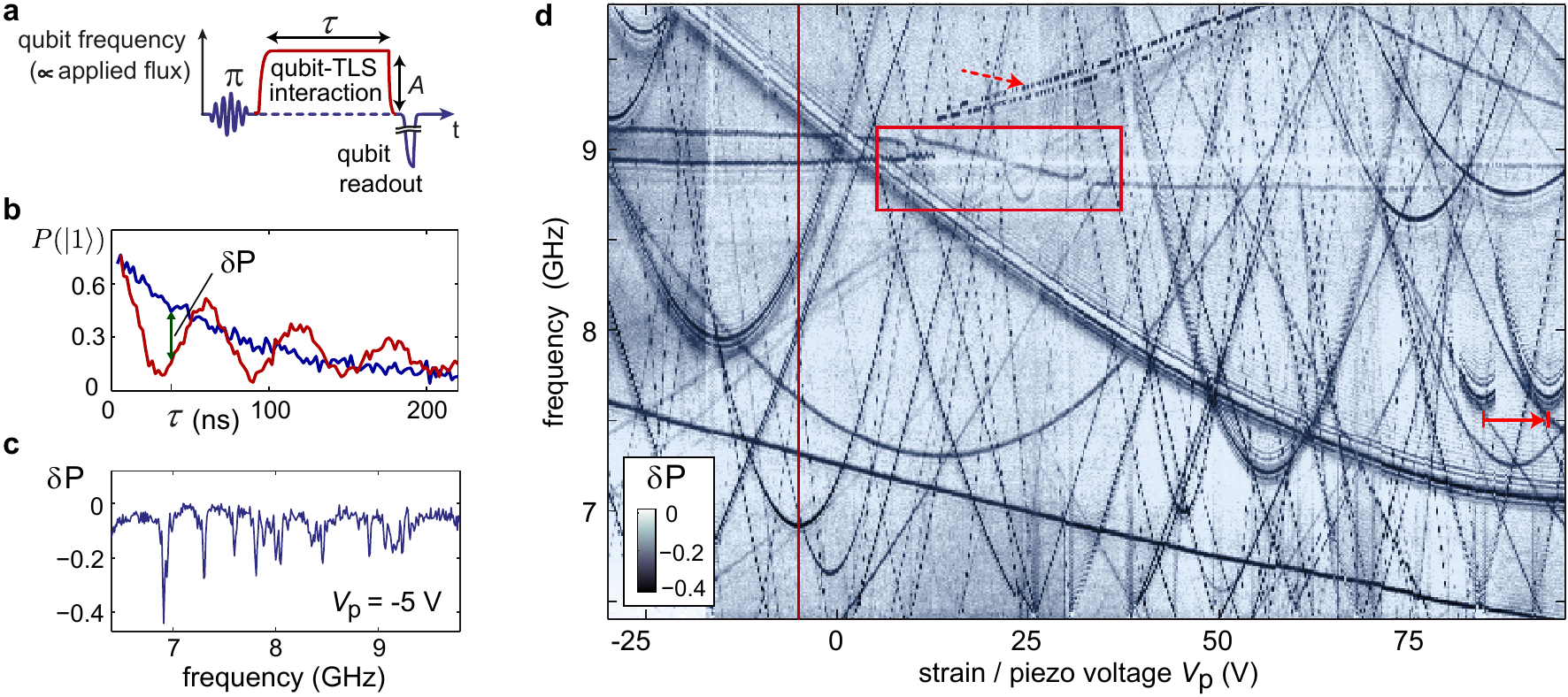}
	\caption{\label{fig:figure2}\textbf{Defect spectroscopy.} \textbf{a} Pulse sequence used to detect TLS via resonant interaction with the qubit. The qubit is excited by a $\pi$ pulse, tuned to varying probe frequencies using different flux-pulse amplitudes $A$, and its population is measured after an interaction time $\tau$. \textbf{b} Qubit population probability, measured for two different probing frequencies in dependence of the interaction time $\tau$. The blue curve shows purely exponential energy relaxation for the isolated qubit, the red curve displays oscillations due to a strongly coupled and coherent TLS. We take the difference between the curves, measured at a fixed $\tau$ as indicated, for the defect signal $\delta P$.
	\textbf{c} Defect signal $\delta P$ in dependence of the probing frequency for a fixed $\tau$. Individual TLS appear as pronounced dips. \textbf{d} Strain-dependence of TLS resonance frequencies, appearing as dark traces in $\delta P$ indicating a reduction in qubit population due to its resonant coupling to a TLS. Mutual TLS interactions are observed as random switching of the TLS resonance frequency (arrows), avoided level-crossings and non-hyperbolic traces (box). The cross-section at $V_\mathrm{p}$=-5~V (vertical line) is shown in \textbf{c}.} 
\end{figure*}

\section*{Results}
\section*{Atomic tunneling systems}
To explain the microscopic origin of the TLS in superconducting electronics, several models\cite{Martin:PRL:2005,Constantin:PRL:2007,DeSousa:PRB:2009,
DuBois:PRL:2013,Agarwal:ArXiv:2012} have been proposed. 
However, all experimental results obtained so far, including the recent demonstration that the energy of the TLS is tunable by static mechanical strain\cite{Grabovskij:S:2012}, are readily explained assuming that they originate from atomic tunneling systems.
As in the well studied model describing the low temperature thermal, dielectric, and acoustic properties of disordered solids\cite{Phillips:JLTP:1972,Anderson:PM:1972,Phillips87}, it is assumed that some atoms or small groups of atoms are able to tunnel between two energetically almost equivalent sites within the disordered oxide material of the device. These systems give rise to two-level excitations in a wide energy range of up to the order $E\approx k_B\cdot$1K or $ \approx h\!\cdot\! 20$ GHz. 
In bulk disordered solids, TLS are found in large numbers but, in contrast to their counterparts present in superconducting qubits, cannot be addressed individually. 

According to the tunneling model, an atomic tunneling system is described as a particle in a double-well potential as shown in Fig.~\ref{fig:figure1}a. 
The energies of the two wells differ by the asymmetry $\varepsilon$ and the tunneling amplitude between them is denoted as $\mathit\Delta$, resulting in a level splitting of the two eigenstates given by $E=\sqrt{\mathit{\Delta}^{2} + \varepsilon^{2} }$.
Tunneling systems couple to the environment predominantly by variation $\delta\varepsilon$ of their asymmetry energy with $\delta\varepsilon$ depending linearly on strain fields and, if the tunneling entity moves a non-zero charge, as well on electric field -- the latter serving as an apparent explanation for the observed coupling of the TLS to the qubit circuit.
A variation $\delta\!{\mathit\Delta}$ of the tunneling amplitude induced by strain or external fields is generally believed to be negligible\cite{Phillips87,Black:PRB:1977}.
We have recently verified the linear strain dependence of $\varepsilon$ and the corresponding hyperbolic variation of the energy splitting $E$ by tracking individual TLS with a phase qubit while bending the chip circuit with a piezo actuator\cite{Grabovskij:S:2012}. In our a setup, sketched in Fig.~\ref{fig:figure1}b, an applied piezo voltage $V_\mathrm{p}$ results in variable strain fields on the order of $10^{-6}/V$.\\

\section*{Defect spectroscopy}
In this study, we detect and analyse TLS using a superconducting qubit. These devices rely on Josephson junctions (JJ) as nonlinear circuit elements, which are realised as two superconducting films separated by a thin, insulating tunneling barrier, consisting of a 2-3~nm thick structurally disordered layer of aluminum oxide. A sketch of the employed phase qubit\cite{Steffen:PRL:2006} including measurement and manipulation circuitry is shown in Fig.~\ref{fig:figure1}c. The qubit's level splitting and their population are controlled by externally applied flux bias and resonant microwave pulses, respectively, and a DC-SQUID is used for qubit readout.

In order to trace the energies of individual TLS while applying strain to the qubit chip, we use a spectroscopy scheme based on the pulse protocol depicted in Fig.~\ref{fig:figure2}a. The qubit is first biased at a frequency far away from the intended spectroscopy region and excited by a resonant microwave $\pi$-pulse.
Applying appropriate flux bias, it is then tuned to the probing frequency $f_{h}$ where it resides for the holding time $\tau$. If at this frequency the qubit is in resonance with a certain TLS, the excitation is shared between the systems\cite{Neeley:NP:2008,Shalibo:PRL:2010}. This results in coherent oscillations which effectively swap the quantum states of the two systems at a frequency determined by their coupling strength as shown in Fig.~\ref{fig:figure2}b. 
A change $\delta P$ in the qubit excitation probability, measured after the interaction time $\tau$, thus reveals the presence of a TLS. We chose $\tau$ to be about half the qubit's life and coherence times $T_1$ and $T_2$ (here, both $\approx 100$ ns) to reach a compromise between the loss of signal due to qubit relaxation and the sensitivity needed for detecting also weakly coupled TLS.
We repeat this procedure for a range of probing frequencies $f_{h}$ and vary the mechanical strain applied to the sample. 

\begin{figure}
\centering\includegraphics{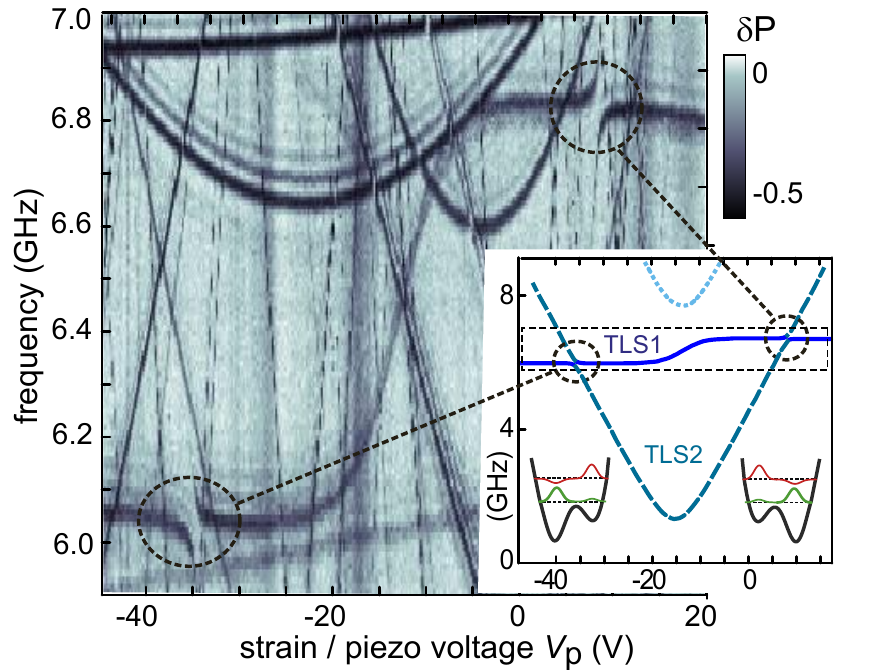}
\caption{\label{fig:figure3}\textbf{Spectroscopic signature of two mutually interacting TLS.} The observed \textit{S}-shaped feature with avoided level crossings (encircled) is characteristic for a TLS (TLS1) that is coherently coupled to a second defect (TLS2). The inset shows their spectrum calculated from theory, closely reproducing the data obtained between 6 and 7 GHz (dashed box). Here, TLS2 is strain-tuned through its symmetry point $\varepsilon_2=0$ and changes the location of the ground state in its double-well potential (insets), while TLS1 is only weakly influenced by strain.} 
\end{figure}

By plotting the change in qubit population as a function of both applied strain and probing frequency, we obtain defect spectra like the one shown in Fig.~\ref{fig:figure2}d. Dark traces indicate the resonance frequencies of individual TLS, which are tuned by strain as expected for atomic tunneling systems. 
Some TLS have a tunneling energy $\it{\Delta}$ that falls within the frequency range accessible by the qubit (about 6.5 - 10 GHz for our sample), while also their asymmetry energy $\varepsilon$ is tuned through zero in the investigated strain range. Accordingly, for those TLS we can clearly observe the hyperbolic strain-dependence of their resonance frequencies around minima given by $\it{\Delta}/h$.
We note that the distribution of TLS resonances changes completely once the sample is warmed to room temperature, see Supplementary Figure 1 for various examples. This can be explained by a modification of the atomic configuration changing the TLS environment locally, and also by an offset in the applied strain due to thermal dilatation of the sample fixture. As long as the temperature is kept below about 10 - 20 K, the properties of the majority of TLS remain constant over several months of measurements.

For TLS that are strongly coupled to the qubit, the chosen interaction time $\tau$ may exceed the duration of one swap operation. Since the latter also depends on the detuning, a fringe-like interference pattern occurs around the traces, see Supplementary Figure 2. More details on this effect and other artefacts in such defect spectra are discussed in Supplementary Note 1. \\

\section*{Mutually coupled TLS}
In the defect spectroscopy example shown in Fig.~\ref{fig:figure2}d, two different effects of interactions between TLS can be identified: i) The resonance frequency of certain TLS is observed to switch between two values in a random, telegraph-signal like fashion, with a switching frequency that depends on the applied strain. This effect is explained by assuming that the observed TLS couples non-resonantly to an incoherent defect which fluctuates between its positions. ii) Much less frequently observed are non-hyperbolic strain responses in the TLS resonance frequency as well as level splittings that are the typical signature of two resonantly coupled coherent quantum systems.

In the remainder of this article, we discuss a particularly clear manifestation of a system of two coupled TLS, whose defect spectroscopy signature revealed the \textit{S}--shaped trace with avoided level crossings shown in Fig.~\ref{fig:figure3}.
To explain this spectroscopic feature, we construct a model of two interacting defects denoted as "TLS1" and "TLS2"  (see also inset to Fig.~\ref{fig:figure3}).
Qualitatively speaking, we observe that TLS1 couples only weakly to the applied external strain so that the step-like increase of its resonance frequency is almost exclusively due to its non-resonant coupling to TLS2. In the region shown, the double-well potential of TLS2 is strain-tuned through its symmetry point $\varepsilon_2(V_\mathrm{p})=0$ at $V_\mathrm{p} \approx -14 V$ such that the probability density of its ground state is shifted gradually from one potential well to the other. The accompanied shift in atomic positions is mediated via internal strain or electric field to TLS1, which responds by modifying its energy splitting, here from 6 to 6.8 GHz. Eventually, the applied strain tunes TLS2 into resonance with TLS1 and their coherent interaction gives rise to the associated and observed level repulsions.

\begin{figure}[htb!]
\centering\includegraphics{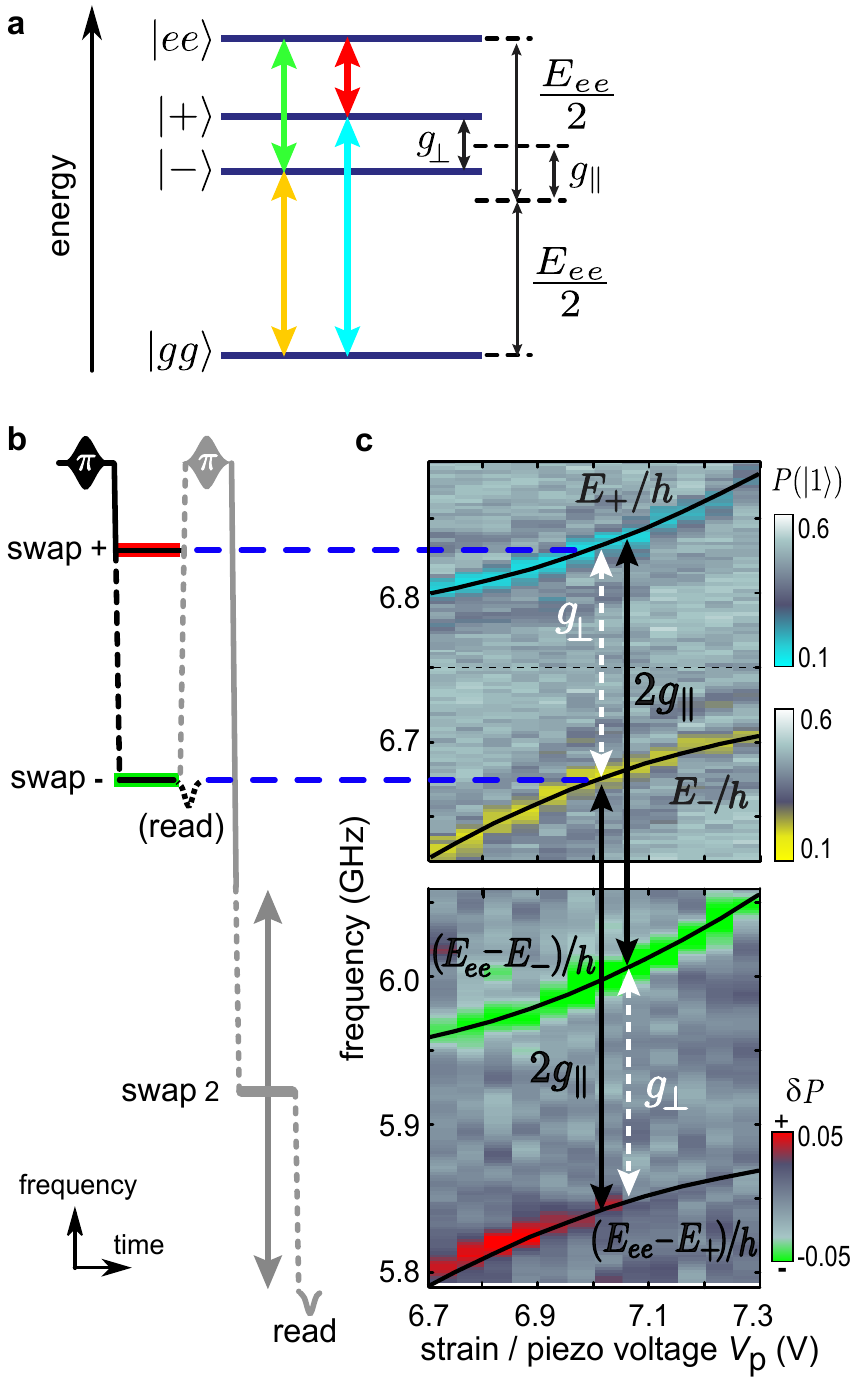}
\caption{\label{fig:figure4} \textbf{Exploring the energy level structure for the coupled defect system.} \textbf{a} Level scheme of the resonantly coupled defect system for an applied strain near the right avoided level crossing of Fig.~\ref{fig:figure3} at $V_\mathrm{p} \approx 7 V$. Energy splitting and offset of the entangled states $|\pm\rangle$ are determined by transversal and longitudinal coupling strengths $g_\perp$ and $g_\parallel$, respectively.
\textbf{b} Pulse sequence used to map out the complete energy level structure. For the upper panel in \textbf{c}, the qubit was measured after the first swap operation at varying frequencies in order to calibrate the entangled states energies. The lower panel was obtained using the complete sequence: preparation of one of the entangled states $|-\rangle$ or $|+\rangle$ in a first swap, followed by a second qubit excitation and variation of the second swap frequency to reveal the transitions between the entangled states to the fully excited state $|\rm{ee}\rangle$. The colour maps were chosen according to the coloured arrows which indicate the corresponding transitions in \textbf{a}.
Lines in \textbf{c} are calculated from theory. 
}
\end{figure}

In the following, we outline our theory of the coupled defect system and show how it can be fully characterised by analysing strain-spectroscopic data.
The Hamiltonian of a single TLS is written as 
\begin{equation}
H_{i} = \frac 12 \varepsilon_i(V_\mathrm{p})\sigma_{\rm{z},i} + \frac 12 {\it\Delta}_i\sigma_{\rm{x},i} = \frac 12 E_i(V_\mathrm{p})\,	\tilde{\sigma}_{\rm{z},i},
\label{eq:H_TLS}
\end{equation}
where ${\it\Delta}_i$ is the tunneling energy and the asymmetry energy $\varepsilon_i(V_\mathrm{p})$ depends linearly on external strain, i.e. voltage $V_\mathrm{p}$ of the piezo drive. Here and in the following, we use the tilde to distinguish operators such as the Pauli-matrices  $\tilde{\sigma}_j$ in the eigenbasis from those in the localized basis $\sigma_j$. 
The energy splitting in the diagonal basis is $E_{i}(V_\mathrm{p}) = \sqrt{\varepsilon_{i}^{2}(V_\mathrm{p}) + {\it\Delta}_{i}^{2}}$. The transformation to the diagonal basis corresponds to a rotation about the angle $\xi_{i}(V_\mathrm{p})$, defined by
$\tan{\xi_{i}}={\it\Delta}_{i}/\varepsilon_{i}$ with
$\sin{\xi_{i}}={\it\Delta}_{i}/E_{i}$ and $\cos{\xi_{i}}=\varepsilon_{i}/E_{i}$. For example, the operators $\sigma_{\rm{z},i}$, whose eigenvalues identify the particle positions, transform as
\begin{equation}
	\sigma_{\rm{z},i} \rightarrow \cos{\xi_i}\,\tilde{\sigma}_{\rm{z},i} +
\sin{\xi_i}\:\tilde{\sigma}_{\rm{x},i}.
	\label{eq:sigma_z-trafo}
\end{equation}

We write the Hamiltionian of two coupled TLS as $H_{\rm{T}}=H_1+H_2+H_{12}$,
with the interaction term
\begin{equation}
	H_{12} = \frac12\,g\,\sigma_{\rm{z},1}\,\sigma_{\rm{z},2} .
	\label{eq:H_12}
\end{equation}

Within the tunneling model, the defect's mutual coupling parameter $g$ comprises their electric dipole interaction as well as a strain-mediated elastic contribution. Our spectroscopic data does not allow us to distinguish between these coupling mechanisms. However, since TLS interact with the qubit only electrically, we can determine the projections of the TLS's electric dipole moments onto the field in the qubit junction independently. Details of this analysis are included in the Supplementary Note 2. 

In the diagonal basis, the interaction Eq.~(\ref{eq:H_12}) consists of four terms, which are combinations of $\tilde{\sigma}_{\rm{z}}$ and $\tilde{\sigma}_{\rm{x}}$ for each TLS (see Supplementary Note 3). However, for explaining the observed S-shaped signal (Fig.~\ref{fig:figure3}), only two terms are relevant. 
The energy shift of TLS1, i.e. the amplitude of the \textit{''S''}, is given by the longitudinal coupling component
$g_{\parallel}=2g \cos{\xi_1} \cos{\xi_2}$ which stems from the term $\propto \tilde{\sigma}_{\rm{z},1}\tilde{\sigma}_{\rm{z},2}$.
The transversal component results in exchange coupling $\propto g_{\perp}\tilde{\sigma}_{\rm{x},1}\tilde{\sigma}_{\rm{x},2}$
and defines the size of the level repulsion 
$g_{\perp}=g \sin{\xi_1}\sin{\xi_2}$.
The  remaining two parts of $\widetilde{H}_{12}$, $\propto\tilde{\sigma}_{\rm{z},1}\tilde{\sigma}_{\rm{x},2}$ and $\propto\tilde{\sigma}_{\rm{x},1}\tilde{\sigma}_{\rm{z},2}$, yield only minor energy shifts and can be neglected to first order. However, in our numerical fits to the spectrum we take the full interaction Eq.~(\ref{eq:H_12}) into account.\\

\section*{Two-photon swap spectroscopy}
To fully explore the coherently coupled defect system with even higher precision, we focus on the region near the right anti-crossing where both TLS are strain-tuned into resonance. This results in the four-level energy spectrum illustrated in Figure~\ref{fig:figure4}a, which we map out by performing microwave swap spectroscopy on the system prepared in different entangled states.
For this, we follow the sequence sketched in Fig.~\ref{fig:figure4}b. The qubit is first prepared in its excited state and then tuned to either the lower or upper branch of the avoided level crossing, realizing a swap operation with the entangled state $\ket-$ or $\ket+$, respectively (see ~\ref{fig:figure4}a). Directly afterwards, the JJ-qubit is again excited and tuned through a lower frequency range in order to find the transition that brings the system of coupled TLS to the fully excited state $\ket{\rm{ee}}$.

Data in the upper panel of Fig.~\ref{fig:figure4}c were obtained by measuring the qubit after the first swap operation, thus indicating the avoided level crossing as in our usual defect spectroscopy protocol in order to calibrate the preparation of the chosen entangled state. The complete sequence results in the data shown in the lower panel of Fig.~\ref{fig:figure4}c, where we plot the difference in qubit population between two experiments in which the TLS system was prepared in either one of the two entangled state by adjusting the frequency of the first swap pulse. 
This experiment clearly reveals the transition to the fully excited state in excellent agreement with theory (dashed lines in Fig.~\ref{fig:figure4}c). Moreover, we directly obtain the energies $E_1$ and $E_2$ of the two unperturbed TLS as well as the longitudinal and transversal coupling strengths as indicated in Fig.~\ref{fig:figure4}a. A detailed analysis of this experiment is contained in Supplementary Note 4.

The spectrum calculated from our theoretical model, shown in the inset to Fig.~\ref{fig:figure3}, allows one to fit all system parameters and reproduces the data with high accuracy (see Supplementary Fig.~3). 
The inter-TLS coupling strength $g = -872$ MHz is calculated from its components $|g_{\perp}| = 155$ MHz and $g_{\parallel}=-428$ MHz, obtained in swap spectroscopy for the TLS tuned into resonance at $V_\mathrm{p}=7$ V as explained above.

Together with the energy splitting $E_i$ of the individual TLS at their resonance, we can determine their strain-dependent mixing angles $\xi_i$ to fully characterize the system.
The obtained TLS parameters including their dependence on applied piezo voltage are summarized in Table~\ref{tab:values},
more details on this evaluation are given in Supplementary Note 3.

\begin{table}
	\begin{tabular}{|c|c|c|}
	\hline  & $\it\Delta_i$ & $\varepsilon_i(V_\mathrm{p})$ @ $\varepsilon_2=0$ \\
 	\hline TLS1 & 5.47 GHz & 3.18 GHz $-$ 4 MHz/V\\
	\hline TLS2 & 1.3 GHz & 295 MHz/V\\
	\hline
	\end{tabular}	
\caption{\label{tab:values} \textbf{Measured parameters of the two coupled TLS.} The asymmetry energies $\varepsilon_i$ are given at the symmetry point of TLS2 ($V_\mathrm{p}=-14.05$~V).}
\end{table}

A final remark concerns the probability of finding two TLS spaced closely enough to expect an interaction of similar strength as in our experiment. Considering only electric dipole interaction and assuming for both TLS an electric dipole moment of $d~\sim 1$\,e\AA, which is consistent with the results of our work and agrees with recent observations and theory\cite{Cole:APL:2010,DuBois:PRL:2013,Martinis:PRL:2005,Agarwal:ArXiv:2012},
we can make a  rough estimate of the maximal distance between the two interacting TLS to be on the order of 5 nm (see Supplementary Note 4). 
When distributing the $\sim 50$ TLS visible in the accessible frequency range of about 6 to 9 GHz evenly onto the area of the $1~\mu$m$^2$ large qubit junction, on average each TLS occupies an area of about 140x140 nm$^2$. Thus, although one may expect an increased TLS density at interfaces, observing two coherently interacting TLS in Josephson junctions is indeed a rare case.\\

\section*{Discussion}
The experimental techniques presented here provide a novel spectroscopic view onto the bath of sparse material defects by accessing the quantum states of individual TLS and small coupled systems with a superconducting qubit while strain-tuning their internal degree of freedom.
This lays the ground for further experiments to clarify the microscopic origin of TLS, which is vital for the advancement of various kinds of nanofabricated devices whose functionality is hampered by defects. 
So far, the atomic tunneling model readily explains all effects observed with TLS in tunnel junctions and, in particular, the here demonstrated strong coupling of TLS to mechanical strain and their interaction with both coherent as well as randomly fluctuating defects. Our results open way to detailed testing of the 50-year old tunneling model on the basis of individual TLS, for example by performing defect spectroscopy for statistical analyses of the TLS distribution and by studying the strain dependence of TLS coherence.

\section{Methods: Superconducting qubit sample}
The phase qubit sample\cite{Steffen:PRL:2006} used in this work was fabricated in the group of J.~M. Martinis at University of California, Santa Barbara (UCSB). The qubit junction had an area of about $1\mu$m$^2$, fabricated using aluminum as electrode material and its thermally grown oxide as a tunnel barrier. All data have been obtained at a sample temperature of about 35 mK. The mechanical strain was controlled by bending the sample chip with a piezo transducer\cite{Grabovskij:S:2012}.

\section{Acknowledgements}

 We would like to thank M. Ansmann and J.~M. Martinis (UCSB) for providing the sample that we measured in this work. We thank E. Demler, T. DuBois, C. Enss, A. Heimes, M. D. Lukin, J.~E. Mooij, A. Shnirman, and A. W\"urger for fruitful discussions. Jared~H. Cole and Clemens M\"uller acknowledge the support of the Australian 
 Research Council and the RMIT Foundation through an International Research Exchange Fellowship. This work was supported by the Deutsche Forschungsgemeinschaft DFG and the State of Baden-W\"urttemberg through the DFG Center for Functional Nanostructures (CFN) as well as by the EU project SOLID. We acknowledge support by Deutsche Forschungsgemeinschaft and Open Access Publishing Fund of Karlsruhe Institute of Technology.
 
 {\bf Author contributions.} The experiments were conceived by J.L., G.G., G.W. and A.V.U. and performed by J.L. and G.G. Theory and simulations were done by C.M., J.H.C. and G.G. The article was written by J.L., G.G., G.W. and C.M. with contributions from all authors.
 Competing Interests. The authors declare that they have no
competing financial interests.
 Correspondence. Correspondence and requests for materials
should be addressed to A.V.U.~(email: ustinov@kit.edu).




\clearpage

\section*{Supplementary Note 1: Defect spectroscopy}
Our defect spectroscopy protocol relies on detecting the transition of a single microwave photon, initially stored in the qubit, via resonant interaction to individual TLS. In comparison to our previous method~\cite{Grabovskij:S:2012}, where a long microwave pulse was saturating the qubit transition, here we use a $\pi$-pulse to maximize the qubit population probability and thus increases the signal-to-noise ratio. In principle, this method equally detects strongly coupled circuit or environmental resonances, which can however be distinguished by their independence on strain. We did not observe any strain dependence of the qubit parameters nor presumed circuit resonances. We note that a similar, independently developed technique of TLS swap spectroscopy was used in Ref.~\cite{Shalibo:PRL:2010} to obtain statistics on TLS coherence and coupling strength, but without mechanical strain control.\\

\noindent\textbf{Observation of coupled TLS systems}

\noindent We performed defect spectroscopy on the same sample and setup in 12 different cool-downs over a course of two years. While the qubit parameters remained constant, the distribution of TLS resonances changed completely. This is not surprising given the observed strong coupling of TLS to the local mechanical strain, which is expected to vary between cool-downs due to small reconfigurations of atomic positions. However, changes in defect spectroscopy can also be explained by offsets in the externally applied strain due to thermal dilatation affecting the sample holder.

\textit{S}-shaped signatures of coherently coupled defects, shown in Supplementary Figure~\ref{fig:spectra}, were observed in two different cool-downs.

\setcounter{figure}{0}
\renewcommand{\thefigure}{\textbf{\arabic{figure}}} 
\renewcommand{\figurename}{\textbf{Supplementary Figure}}
\renewcommand{\theequation}{\arabic{equation}} 
\begin{figure*}[htb!]
\centering\includegraphics[width=15cm,height=21cm]{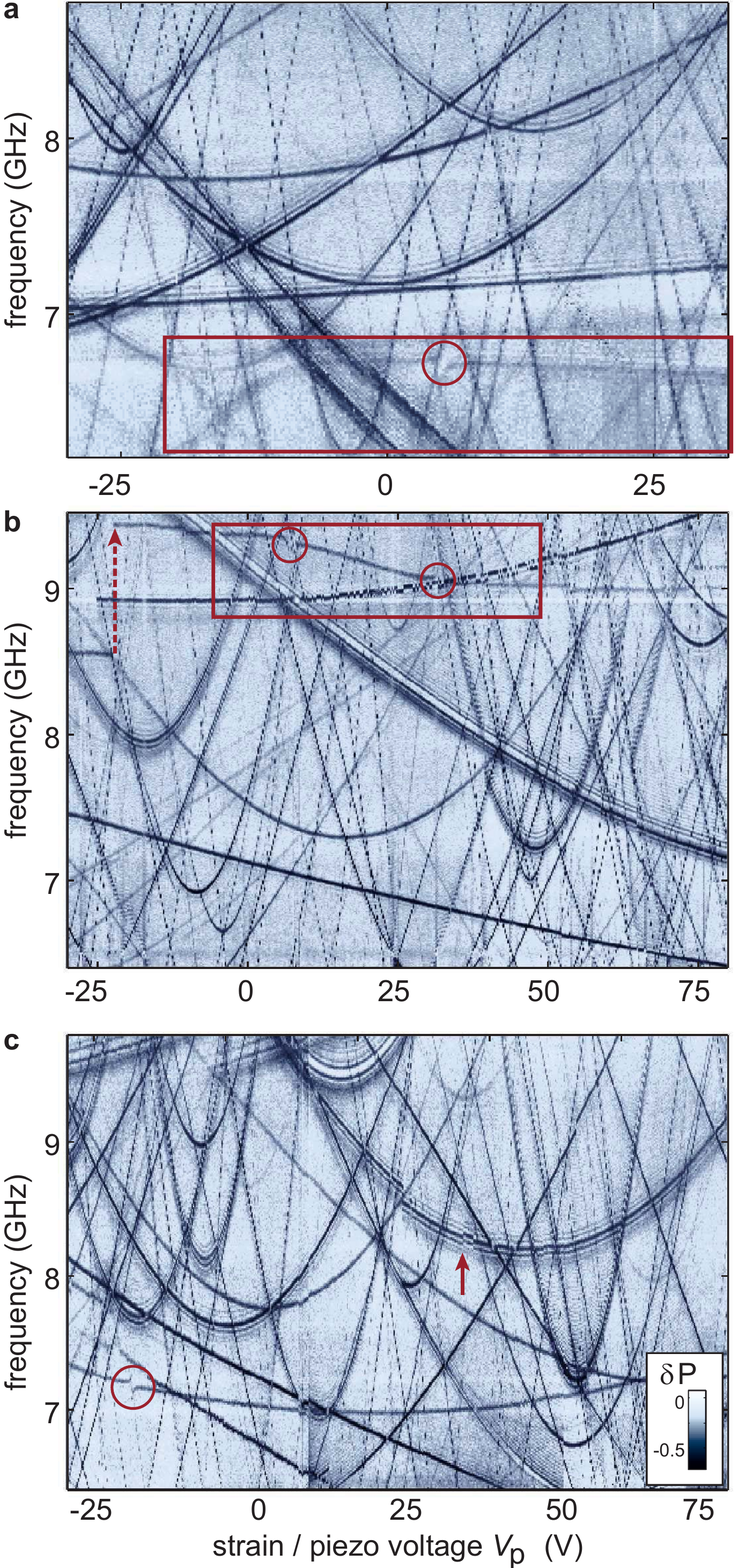}
\caption{\label{fig:spectra} Defect spectroscopy examples obtained on the same sample. Red boxes highlight coupled TLS systems, circles indicate avoided level crossings.
\textbf{a} Data taken in April 2012, showing partly the signature of the coupled TLS system that was characterized in this work.
\textbf{b} Data acquired soon after cooling down the sample in March 2013, about 3 weeks before taking the data in Fig. 2d of the main manuscript. A few TLS changed their properties, while most remained stable. The dashed arrow indicates a spontaneous change in TLS parameters.
\textbf{c} Another example of an observed avoided level crossing (circle), and telegraphic switching (arrow).}
\end{figure*}

\begin{figure*}[htb]
\centering\includegraphics[scale=1.2]{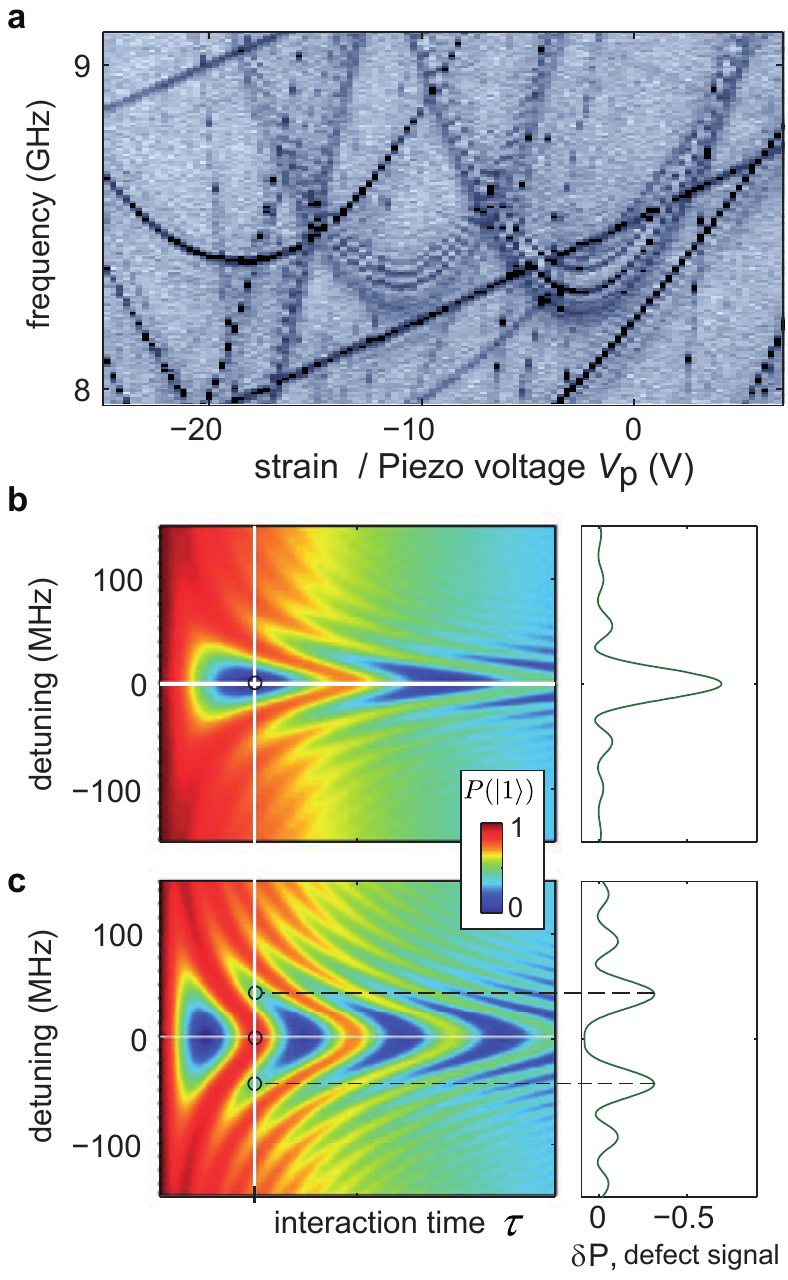}
\caption{\label{fig:fringes}\textbf{Origin of fringes in defect spectra.}\textbf{a} 
Defect spectroscopy example, showing the hyperbolic resonance frequency traces of three TLS. Fringes appear around the exact TLS resonance for strongly coupled TLS (middle and rightmost hyperbola). \textbf{b} and \textbf{c} Calculated population probability $P(|1\rangle)$ of an initially excited qubit, showing coherent oscillations due to coupling to a TLS in dependence of the interaction time~$\tau$ and detuning (left panels). The right panels show the defect signal $\delta P$ for a fixed interaction time $\tau$ (cross-section along vertical white line in the left panel). For \textbf{c}, the TLS-qubit coupling strength is twice larger than for \textbf{b}, resulting in faster oscillation. For the more strongly coupled TLS, but same interaction time $\tau$, the signal $\delta P$ reaches minima at non-zero detuning, giving rise to the pronounced fringes or shadow-lines in defect spectroscopy.
}
\end{figure*}

Supplementary Figure~\ref{fig:spectra}\textbf{a} was obtained in the experimental run in which we characterized the coupled defect system as discussed in the main text. About one year later, the same sample revealed a similar system of two coupled defects shown in the red box in ~\ref{fig:spectra}\textbf{b}, having comparable longitudinal coupling strength but different resonance frequencies and opposite symmetry. It may well be that both observations involve the same tunneling systems which however experience slightly different local potential configurations.

In Supplementary Figure~\ref{fig:spectra}\textbf{b}, one can notice that the frequency of the coupled defect system switched while the spectrum was measured around a voltage of about -25 V (the dashed red arrow indicates the shift in frequency). Such spontaneous changes in the resonance frequencies of some TLS are observed frequently and occur most often shortly after the sample was cooled to milikelvin temperatures. This is readily explained by a coupling of the observed coherent TLS to incoherent TLS, also called "two-level fluctuators", which may be of same physical origin but have a small rate of tunneling between their potential wells. If such a fluctuator is strongly asymmetric, it may be found in its higher energy state soon after the cool-down while it would remain in its ground state once tunneling occurred.
Experimentally, we found that the number of spontaneously switching TLS is reduced by repeatedly cycling the applied strain through the complete range, hereby "annealing" the sample in the sense that fluctuators are stimulated to tunnel to their more stable ground state. The data shown in Fig.~\textbf{2~d} of the main text has been measured in the same cool-down as \ref{fig:spectra}~\textbf{b}, but 3 weeks later. One can see that some TLS changed their properties in between measurements, while others remained stable. 
More examples of mutual defect coupling are seen in Supplementary Figure~\ref{fig:spectra}\textbf{c}, showing telegraphic switching of a TLS' resonance frequency (arrow) and an avoided level crossing (circle).\\

\noindent\textbf{Effects and experimental artefacts}\\
\noindent
In the following, we discuss some artefacts that may be observed with this spectroscopy technique.

Fringes or shadow-like replica of some TLS traces, such as shown in Supplementary Figure~\ref{fig:fringes}\textbf{a}, may appear for TLS whose coupling strength to the qubit exceeds $h/2\tau$. In this case, the chosen interaction time $\tau$ exceeds the duration of one swap operation such that the excitation is transferred back to the qubit, and this lowers the defect signal $\delta P$ (see also Fig.~2b in main article). Since the frequency of swap oscillations increases with the detuning between qubit and TLS, a minimal qubit population (maximal signal $|\delta P|$) thus occurs only in the vicinity of the exact resonance as shown in Supplementary Figures~\ref{fig:fringes}\textbf{b} and \textbf{c}.

Fringes may also occur due to residual entanglement between the qubit and TLS. An example is visible in Supplementary Figure~\ref{fig:spectra}\textbf{c} in a range of $V_\mathrm{p}\approx 40...50$~V at low frequencies. In this measurement, the qubit was biased at a frequency of 6.4 GHz, so that already during excitation it was near resonance with a strongly coupled TLS. Accordingly, the $\pi$-pulse prepares some entangled state between the systems, whose phase is then modified in dependence of the subsequent qubit detuning during the interaction time~$\tau$.

Qubit drift due to uncontrolled slow changes in its bias flux leads to deviations in the defect signal's frequency dependence. 
For Supplementary Figure~\ref{fig:spectra}\textbf{c}, in total about 140,000 data points were measured, of which each was averaged 1000 times at a repetition rate of about 500 Hz, resulting in a total measurement duration of $\approx$ 3.5 days. An example of qubit drift can be seen in Supplementary Figure~\ref{fig:spectra}\textbf{c} around a piezo voltage of about 7 V, which was measured while feeding liquid Helium to the dilution refrigerator. This resulted in a small change of the qubit flux bias because of temperature variations in its bias line filters that are installed at the 1K-pot of the cryostat. Qubit drift can be distinguished from TLS resonance frequency fluctuations since all TLS signals are shifted equally. We note that the measurement time could be easily reduced by at least one order of magnitude if one employs a dispersive qubit readout method~\cite{Wirth2010,Jerger2011,Chen2012} instead of the DC-SQUID switching-current measurements employed here~\cite{wallraff03}.

\section*{Supplementary Note 2: The qubit-TLS coupling}
Assuming the TLSs interact with their environment via their electric dipole moment with the qubit circuit, the corresponding coupling operator is $\sigma_{z}$, whose expectation values identify the position of the particle. The interaction of an individual TLS with the electric fields inside the qubit's junction is then described as
\begin{align}
	H_{iq} = \frac12 v_{i} \tilde{\tau}_{x} \sigma_{z,i} \,,
	\label{eq:hiq}
\end{align}
where $v_{i}$ is the electric dipole interaction strength and we used the fact
that for a phase qubit in its eigenbasis, the electric field operator is 
$\propto\!\!\tilde{\tau}_{x}$~\cite{Cole:APL:2010}. 
Rewriting Supplementary Eq.~(\ref{eq:hiq}) in the TLS eigenbasis, we retain two terms, of which the transversal coupling term $v_{i,\perp}\tilde{\tau}_{x}\tilde{\sigma}_{x,i}$
with $v_{i,\perp}=v_i\sin\xi_i$ describes the resonant interaction between the JJ qubit and the TLS $i$. It is this term in particular which causes the exchange of excitations between the two systems.

We obtained the electric coupling strength between qubit and TLS1, $v_{1,\perp}$, by fitting time-domain oscillations of the initially excited qubit tuned into resonance with TLS1, far away from the TLS1-TLS2 anti-crossing, where $v_{1,\perp}$ is independent of the state of TLS2. We performed the same experiment also with both, lower and upper branches, in the vicinity of the TLS1-TLS2 anti-crossing (the pulse sequence is shown on the inset in Supplementary Figure~\ref{fig:figure4_app}). The result is plotted in Supplementary Figure~\ref{fig:figure4_app}\textbf{a}, green for the lower and red for the upper branch.
The panels in Supplementary Figure~\ref{fig:figure4_app}b and c show the observed oscillations in the qubit population.

Since TLS2 is very weakly coupled to the qubit, the oscillation frequency decreases quickly with voltage when following a branch turning into TLS2. However, it is worth noting that the crossing point of the two curves is not at the voltage of 7 V where TLS1 and TLS2 are in exact resonance. This would be expected if the qubit-TLS2 coupling was zero, as indicated by the dashed curves. A non-zero $v_{2,\perp}$ explains this shift, because the coupling between the qubit and the $\ket{\pm}$ states scales as $(v_{1,\perp}\pm v_{2,\perp})/\sqrt{2}$. Then, the theory shows a very good agreement with the experiment and we obtain $v_{1,\perp}= 15.4$ MHz and $v_{2,\perp}= 3.0$ MHz at an applied strain where the TLS are in resonance $E_1 = E_2$.

With the already known $\xi_1$ and $\xi_2$, $v_1 = 18.7$ MHz and $v_2=14$ MHz can be calculated from Supplementary Equation~(\ref{eq:Ham_full_param}). It is worth mentioning that the very small $v_{2,\perp}$ is not attributed to a smaller dipole moment in direction of the qubit electric field, but can be simply explained by a strong asymmetry ${\mathit\Delta}_2 \ll \varepsilon_i$ resulting in a very small dipole moment of the TLS eigenstates. This analysis is restricted to the assumption that both TLSs interact purely electrically.

\begin{figure*}[htb!]
\centering\includegraphics[scale=0.9]{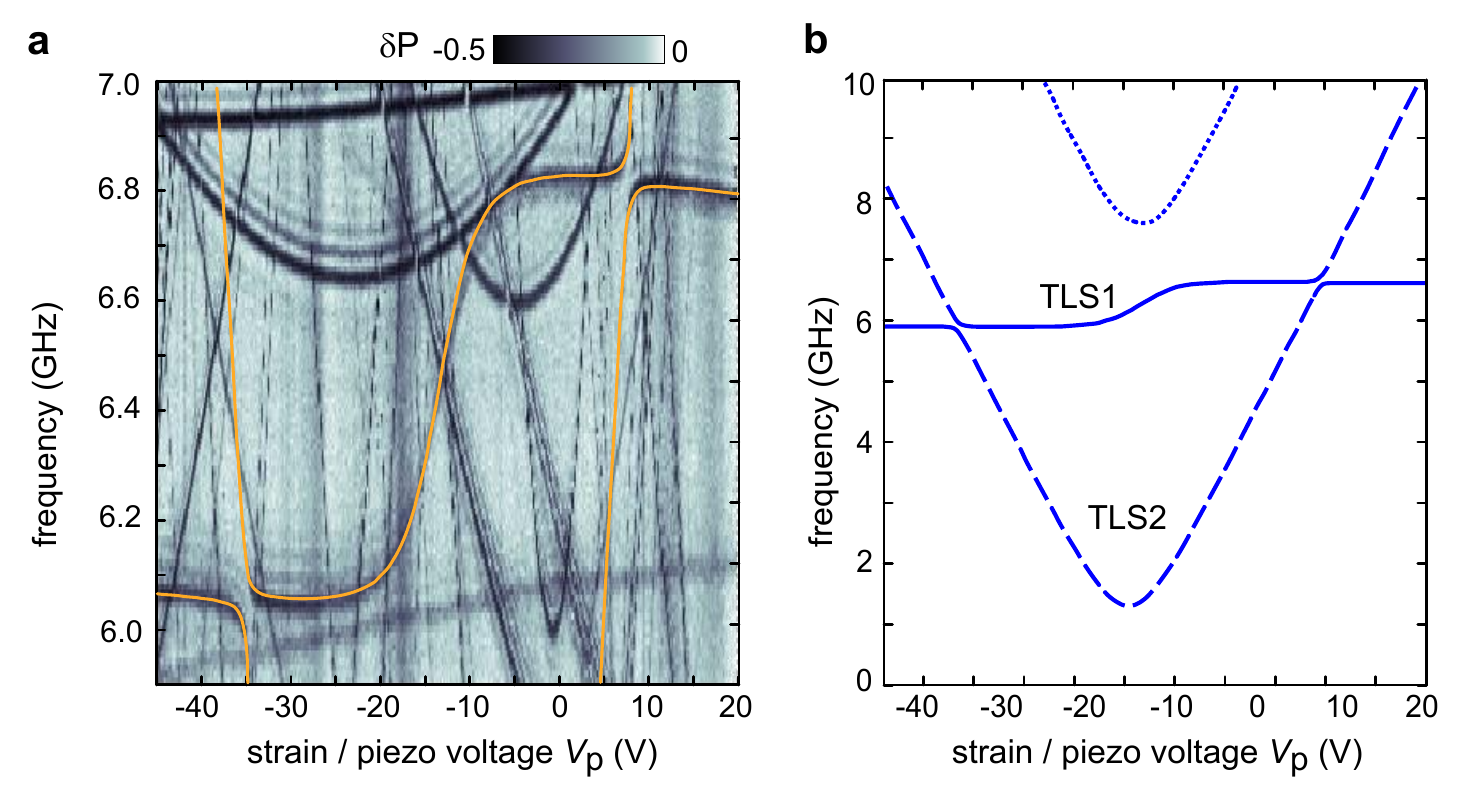}
\caption{\label{fig:figure3_app} \textbf{Comparison between experiment and theory.}
\textbf{a} Strain-spectroscopy data showing the characteristic \textit{S}-shaped signature of two coupled defects. The calculated spectrum is superimposed on the data, showing the excellent agreement with theory. \textbf{b} A plot of the whole theoretical spectrum using the values from Tab.~I. The dashed hyperbola visualizes the TLS2 trace invisible in spectroscopy, while the dotted curve shows the energies of the fully excited system.}
\end{figure*}

\begin{figure*}
\centering\includegraphics[scale=0.9]{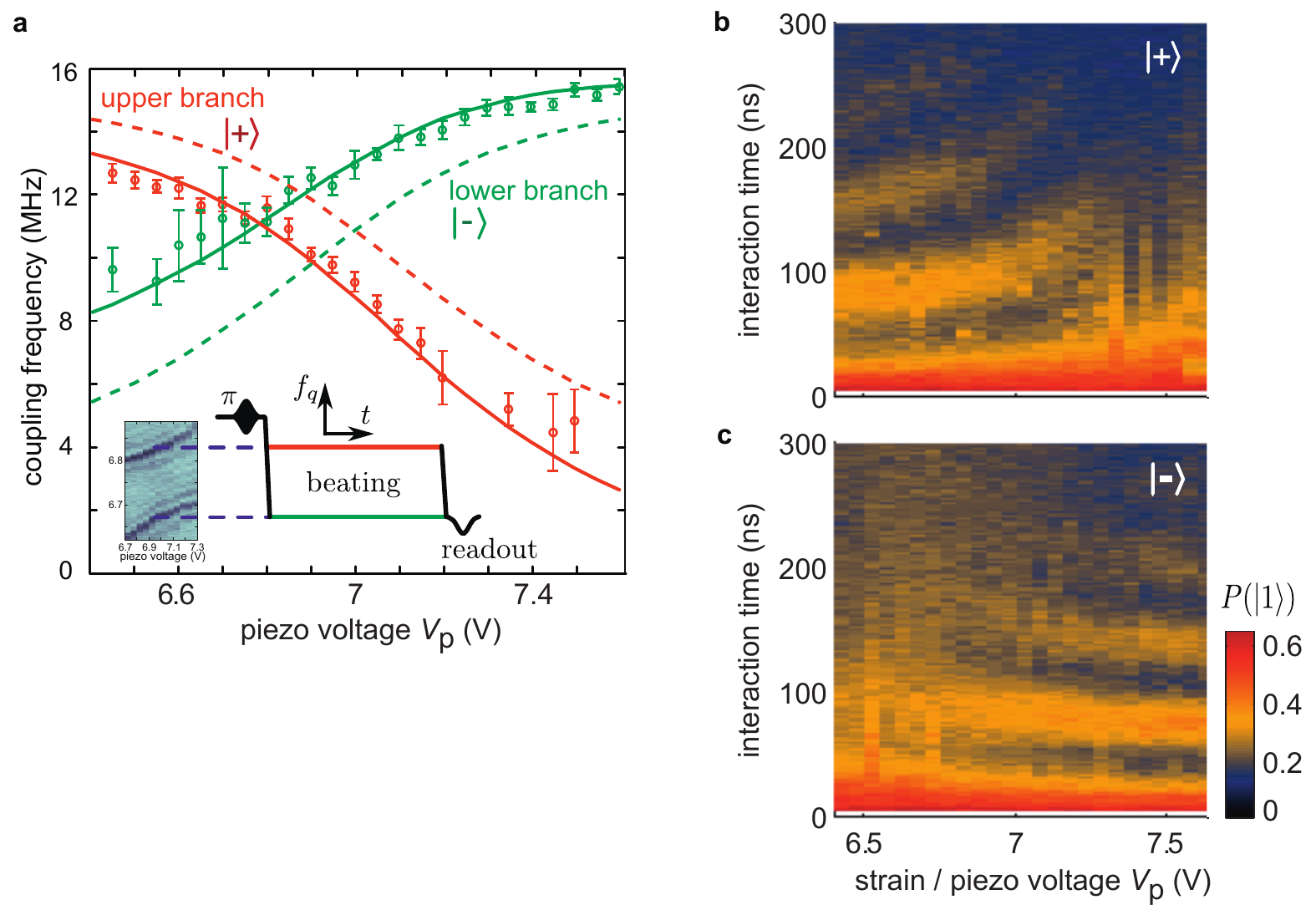}
\caption{\label{fig:figure4_app} \textbf{Measuring the coupling strength between the qubit and each TLS.}
\textbf{a} Measured frequencies of oscillations in the qubit population for the excited qubit tuned to lower and upper branches of the avoided level crossing, respectively, which equals the coupling strength to the corresponding entangled states $|+\rangle$ and $|-\rangle$.
The inset shows the used pulse sequence. It can be clearly seen that the measured coupling strengths (circles) deviate from the dashed curves which are calculated for the case of $v_{2,\perp}=0$. However, one can account for the shift by introducing a non-zero qubit-TLS2 coupling (solid lines). 
\textbf{b} and \textbf{c}: Observed coherent swap oscillations in $P(|1\rangle)$ (colour-coded) vs. strain and the interaction time. Their decay is dominated by qubit decoherence.}
\end{figure*}

\section*{Supplementary Note 3 : Analysis of the full Hamiltonian}
We write the full Hamiltonian describing the JJ-qubit coupled to two
interacting TLSs as  
\begin{align}
	H = H_{q} + \sum_{i=1}^{2} H_{i} + \sum_{i=1}^{2} H_{iq} + H_{12} \,,
	\label{eq:HFullsupp}
\end{align}
with the qubit Hamiltonian $H_{q} = \frac12 E_{q} \tilde{\tau}_{z}$, where $E_{q}$ is the level-splitting of the two lowest qubit states and $\tilde{\tau}_{z}$ is a Pauli-matrix describing the qubit eigenstates. $H_{i}$ is the Hamiltonian of the $i^{\rm th}$ TLS, while $H_{iq}$ describes its interaction with the qubit, and $H_{12}$ the TLSs' mutual interaction as described in the main manuscript.

Diagonalizing the Hamiltonian in Supplementary Eq.~(\ref{eq:HFullsupp}) introduces several new terms. The resulting Hamiltonian, $\widetilde{H}$, can be significantly simplified by ignoring all coupling terms of the form $\propto \sigma_x\sigma_z$ and $\propto \sigma_z\sigma_x$. 
They represent a coupling where one partner changes its state depending on the instantaneous state of the other subsystem and contribute only as small energy offsets.
Neglecting these minor energy shifts, we can write the significant parts of the full Hamiltonian as
\begin{align}
	\widetilde{H}&= \frac{1}{2} (E_q\tilde{\tau}_z + E_1\tilde{\sigma}_{z,1}+ 
	E_2\tilde{\sigma}_{z,2}+g_\parallel\tilde{\sigma}_{z,1}\tilde{\sigma}_{z,2}\nn\\
	&+g_\perp\tilde{\sigma}_{x,1}\tilde{\sigma}_{x,2}+v_{1,\perp}\tilde{\sigma}_{x,1}\tilde{\tau}_{x}+v_{2,\perp}\tilde{\sigma}_{x,2}\tilde{\tau}_{x}),
\end{align}
with
\begin{align}
	E_i = \sqrt{\varepsilon_i^2+{\it\Delta}_i^2} &\,,\nn\\
	v_{1,\perp} = v_1\sin\xi_1  \quad&\,,\quad  v_{2,\perp} = v_2\sin\xi_2 \,,\nn\\ 
	g_\parallel = g\cos\xi_1\cos\xi_2  \quad&\,,\quad g_\perp = g\sin\xi_1\sin\xi_2 \,,\nn\\
	\cos{\xi_{i}} = \varepsilon_{i}/E_{i} \quad&\,,\quad \sin{\xi_{i}}={\it\Delta}_{i}/E_{i}	
	\label{eq:Ham_full_param}	
\end{align}
Here the relationship between the longitudinal and transversal coupling factors $g_{\parallel}$ and $g_{\perp}$, which are easily identifiable in experiment, and the TLS mixing angles $\xi_{i}$ becomes clear. In our case, the $\xi_i$ dependence of $g_\perp$ cannot be observed because when TLS2 is detuned from TLS1, the perpendicular coupling $\propto \tilde\sigma_{x,1}\tilde\sigma_{x,2}$ becomes irrelevant 
We can interpret the term $\cos\xi_i$ as representing the expectation value of the position operator in the double-well potential $\mean{\sigma_{z,i}}$, so that $g_\parallel=g\mean{\sigma_{z,1}}\mean{\sigma_{z,2}}$. As explained in the main manuscript, when tuning through the TLS2 symmetry point, $\mean{\sigma_{z,2}}$ changes its sign, corresponding to reversing the direction of TLS2's dipole moment. Due to the strong dipole-dipole interaction, TLS1 adjusts its energy splitting accordingly, and this results in the observed \textit{S}-shaped signature.

Additionally, the angles $\xi_i$ also change the qubit-TLS coupling strength $v_{i,\perp}$ which determines the visibility of TLS in our defect spectroscopy. The coupling strength $v_{i,\perp}$ is maximal when the TLS is at its symmetry point $\varepsilon_i=0$ , and it is strongly suppressed for largely asymmetric TLSs with $\it\Delta_i \ll \varepsilon_i$. These considerations allow us to obtain a close fit of the theoretical model to the experimental data, with the results displayed in Supplementary Figure~\ref{fig:figure3_app} and table~I, respectively. 

During our measurements, we occasionally observe abrupt jumps of individual TLS' resonance frequencies on a time-scale of days or weeks. From defect spectroscopic data obtained before and after such events, we found that the defects experience changes of both their asymmetry and tunneling energies, while the strain dependence of $\varepsilon$ remains very similar. A sudden change in TLS parameters also occurred between measurements of the defect spectrum and that of the fully excited TLS1-TLS2 system (see Appendix B), by which the TLS1 energy shifted by approximately 100~MHz towards lower frequencies. Since after the jump, the theoretical spectrum is found to still coincide with the anti-crossing on the left side of the ``S'' (data not shown), we can conclude that ${\it\Delta}_2$ did not change appreciable in comparison to $\varepsilon_2$ at the TLS1-TLS2 anti-crossing.

\section*{Supplementary Note 4: Measuring the energy of the fully excited TLS1-TLS2 system}

The aim of the experiment discussed here is to measure the transition energies of the coupled TLS1-TLS2 system around the anti-crossing at $V_\mathrm{p}=7$~V on the right side of the ``S''-shaped signature shown in Supplementary Figure~\ref{fig:figure3_app}. The energies of the four levels $\ket{gg}$, $\ket{\pm}$ and $\ket{\rm{ee}}$ are illustrated in Fig.~4~a. While the splitting size equals $g_\perp$, the energetic shift of the anti-crossing from $E_{\rm{ee}}/2$, which corresponds to the unperturbed TLS energies $E_1$ and $E_2$, yields $g_\parallel$~\cite{Cole:APL:2010,Lupascu:PRB:2008}. Fig.~4~b visualizes the pulse sequence for this experiment. The upper image of Fig.~4~c shows a zoom of the right anti-crossing with the two branches. After exciting the qubit with a $\pi$-pulse, the excitation is swapped to one of the branches by tuning the qubit for the swap time to the corresponding energies around $E_+$ or $E_-$. Afterwards, the qubit is excited a second time and tuned to lower frequencies in order to search for the transition that fully excites the TLS system. This occurs at energies around $E_{\rm{ee}}-E_+$ or $E_{\rm{ee}}-E_-$, respectively, and the qubit will loose its excitation yielding an additional dark trace in the spectrum image (data not shown). The two plots obtained in this way, one for each branch into which the first excitation was swapped, are subtracted from each other yielding peaks or dips at the relevant energies and zero for the background (Fig.~4~d, lower plot). For better visibility, a color-map has been chosen such that peaks in Fig.~4~c, which correspond to first exciting the lower branch, appear in green and the dips, arising if the qubit excitation was first swapped to the upper branch, show up red. The coupling constants between the two TLSs, and also the TLS energies in the uncoupled case, can now be directly extracted from the data: 
\begin{align}
	2E_1 &= 2E_2 = E_++(E_{\rm{ee}}-E_+) = E_-+(E_{\rm{ee}}-E_-)\nn\\
	g_\perp &= E_+-E_- = (E_{\rm{ee}}-E_-)-(E_{\rm{ee}}-E_+)\nn\\
	2g_\parallel &= E_+-(E_{\rm{ee}}-E_-) = E_--(E_{\rm{ee}}-E_+).
\end{align}
With the known ratio of the tunneling and asymmetry energies of TLS2, $\xi_2$, $g$ and $\xi_1$ (Supplementary Eq.~(\ref{eq:Ham_full_param})) can be calculated (see Tab.~I). 

\section*{Supplementary Note 5: Distance between the TLS}
In order to obtain a rough estimate of the distance between the coupled TLS, several assumptions have to be made. First of all, for the calculation we assume that the mutual TLS interaction is solely due to electrical dipole coupling, i.e. we neglect any contribution due to elastic coupling.
The interaction strength can then be written as~\cite{Kocbach-DipDip-2010}
\begin{align}
\frac{g}{2} = \frac{1}{4\pi\varepsilon_0\varepsilon_rr^3} \left(\vec{\mu}_{1\perp}\vec{\mu}_{2\perp}-2\vec{\mu}_{1\parallel}\vec{\mu}_{2\parallel}\right),
  \label{eq:Hdd-perp-par}
\end{align}
where $\vec{r}$ is the relative position vector from TLS1 to TLS2
and their electrical dipole moments $\vec{\mu}_i=\vec{\mu}_{i,\parallel}+\vec{\mu}_{i,\perp}$
are decomposed into components parallel and perpendicular to $\vec{r}$, respectively.

We estimate the magnitude of the TLS' electrical dipole moments by their components parallel to the electric field in the Josephson junction, which can be determined as described in Supplementary Note 3. The qubit-TLS coupling strength $v_{i,\parallel}$ is the product between the TLS dipole moment and the electrical field in the junction~\cite{Martinis:PRL:2005},
\begin{align}
v_{i,\parallel} = 2 \frac{\mu_{i,\parallel}}{x}\sqrt{\frac{E_{10}}{2C}},
\end{align}
where $x\approx$ 2 nm is the thickness of the Junction's tunnel barrier, $C= 850$ fF the qubit's capacitance and $E_{01} \approx h\cdot 6.5$ GHz the qubit's energy level splitting at resonance with the TLS. Using the values of Table~I, we obtain $\mu_{1,\parallel}\approx 0.46$ e\AA~  and $\mu_{2,\parallel}\approx 0.38$ e\AA. Finally, we assume $\mu_i = \mu_{i,\parallel}$ and use Supplementary Eq.~(\ref{eq:Hdd-perp-par}) with $\epsilon_r=10$ (saphhire) to estimate the maximal distance between the TLS, which results in r = 4.1 nm and r = 5.2 nm, for horizontal and vertical arrangements of the TLSs, respectively. We note that this maximal vertical distance exceeds the thickness of the tunnel barrier.
\clearpage
\section*{Supplementary References}
\vspace{-1cm}


\end{document}